\documentclass[11pt]{article}
\usepackage{graphicx}
\usepackage{moriond}

\def\be{\begin{equation}}
\def\ee{\end{equation}}
\def\bea{\begin{eqnarray}}
\def\eea{\end{eqnarray}}

\def\ppbar{\mbox{p}\overline{\mbox{p}}}
\def\bbbar{\mbox{b}\overline{\mbox{b}}}

\def\Wjets{{\rm W + jets}}
\def\pttau{{p_T^{\tau}}}

\def\MET{{E\kern -0.6em/_{\rm T}}}
\def\METx{{E\kern -0.6em/_{\rm x}}}
\def\METy{{E\kern -0.6em/_{\rm y}}}
\def\PET{{P\kern -0.6em/_{\rm T}}}
\def\Ztt{{\mathrm Z \rightarrow \tau\tau}}

\def\Wmn{{\mathrm W \rightarrow \mu\nu}}

\def\pT{p_{\mathrm T}}

\def\CLS{{\rm CL}_{\mathrm S}}
\begin{document}
\begin{flushright}
FERMILAB-CONF-07-084-E \\
MAN/HEP/2007/9
\end{flushright}
\vspace*{4cm}
\title{Search for Neutral Higgs Boson Production in
the Decay~\boldmath$h\to \tau_{\mu}\tau$  
with the D{\O} Detector\unboldmath \\ \vspace*{2.0cm}}

\author{ Mark Owen \\ On Behalf of the D{\O} Collaboration }

\address{School of Physics and Astronomy, University of Manchester,
Oxford Road, M13 9PL, UK}

\maketitle\abstracts{
A search for the production of neutral Higgs bosons decaying
into $\tau^+\tau^-$ final states is presented. One of
the two $\tau$ leptons is required to decay into a muon.
The data were collected by the
D\O~detector and correspond to an integrated luminosity of
about $1.0$~fb$^{-1}$. 
No excess is observed above the expected backgrounds.
The results are interpreted 
in the Minimal Supersymmetric Standard Model. In
the mass range $90<m_A<200$~GeV values of  $\tan\beta$  larger than 40-60
are excluded for the no-mixing and the $m_h^{\rm max}$ benchmark
scenarios.
}

\section{Introduction}

The contribution of $\tau^+\tau^-$ final states from the Standard Model (SM) Higgs 
production is too small to play any role in the SM Higgs searches
in $\ppbar$ collision at the Tevatron due to the large irreducible background from 
$Z\rightarrow\tau^+\tau^-$ production. 
This is different in the 
Minimal Supersymmetric Standard Model (MSSM), which introduces
two Higgs doublets leading to five
Higgs bosons: a pair of charged Higgs boson (H$^{\pm}$); two neutral
CP-even Higgs bosons (h,H) and a CP-odd Higgs boson (A). 
At tree level, the Higgs sector of the MSSM is fully described
by two parameters, which are chosen to be the mass of the CP-odd
Higgs, $m_A$, and the ratio of the vacuum expectation
values of the two Higgs doublets, $\tan \beta$. The Higgs production cross-section
is enhanced in the region of high $\tan\beta$~\cite{bib-carena}.
In the low $m_A$, high $\tan\beta$ region of the parameter space, 
Tevatron searches can therefore
probe several MSSM benchmark scenarios extending the search regions
covered by LEP~\cite{bib-lep}. 
Inclusive searches for 
$\phi (=H,h,A)\to\tau\tau$ have been performed with integrated 
luminosities of $L=350$~pb$^{-1}$ by D\O\ \cite{bib-d0} and
$L=310$~pb$^{-1}$ by CDF~\cite{bib-cdf}. Both searches require
at least one $\tau$ lepton to decay into an electron $(\tau_e)$ or a muon $(\tau_{\mu})$.
In this analysis, only
the decay $\phi\to\tau_{\mu}\tau$ is considered using
an integrated luminosity of $L=1.0$~fb$^{-1}$.
CDF has also recently released a preliminary result with 
$L=1.0$~fb$^{-1}$~\cite{bib-cdf-1fb}.
The biggest improvement in sensitivity compared to the previous analysis
comes from using a neural network to improve the separation between
signal and background.

\section{Event Preselection}

The preselection requires one isolated muon
with, $\pT^{\mu} > 15$~GeV. 
The event is required to have no other muon
that is matched to a track in the central detector with $\pT^{\mu} > 10$~GeV.

Hadronically decaying taus are characterized by a narrow isolated
jet that is associated with three or less tracks. Three types
of hadronically decaying taus are distinguished:
\begin{itemize}
  \item[{\bf Type 1:}] Calorimeter energy cluster, with one associated track and no
    electromagnetic sub-cluster. This corresponds mainly to the decay 
    $\tau^{\pm} \rightarrow \pi^{\pm} \nu$.
  \item[{\bf Type 2:}] Calorimeter energy cluster, with one associated track and
    at least one electromagnetic sub-cluster. This corresponds mainly to the decay
    $\tau^{\pm} \rightarrow \pi^{\pm} \pi^{0} \nu$.
  \item[{\bf Type 3:}] Calorimeter energy cluster, with 
    three associated tracks,
    with an invariant mass below $1.7$~GeV.
    This corresponds mainly to the decays
    $\tau^{\pm} \rightarrow \pi^{\pm} \pi^{\pm} \pi^{\mp} (\pi^{0}) \nu$.
\end{itemize}
Tau decays into electrons are usually reconstructed as type-2 taus.
These are not removed from the sample.
The event is required to contain a $\tau$ candidate at a distance 
$\Delta R > 0.5$
from the muon direction.
The charge of the $\tau$ candidate
must be opposite to the muon charge.
The transverse momentum $\pttau$ of the $\tau$ candidate measured in the calorimeter must be
greater than $15$~GeV for $\tau$-type 1 and 2, and greater than $20$~GeV 
for $\tau$-type 3. At the same time the transverse momentum of the track
associated with the $\tau$ candidate is required to be $p_T > 15$~GeV for
$\tau$-type 1 and $p_T > 5$~GeV for $\tau$-type 2. 
In the case of $\tau$-type 3, the scalar sum
of the transverse momenta of all associated tracks must be 
greater than $15$~GeV.

\section{\boldmath$\Wmn$ and Multi-jet Background Estimation}

The shape of the $\mathrm{W}\rightarrow\mu\nu + \mathrm{jet}$ background distribution, 
where the jet is misidentified as a tau, was simulated using PYTHIA. 
The normalization, however, was obtained using data. 

A contribution to the background is expected from heavy flavour multi-jet events,
where a muon from a semi-leptonic decay passes the isolation 
requirement and a jet is mis-identified as a $\tau$ candidate.
In addition, a contribution is expected from light quark multi-jet events
where the jets fake both the tau and the muon.
The multi-jet background shape is taken from 
events with at least one muon and one $\tau$ candidate 
where the muon failed the calorimeter
isolation requirement. 
The normalization of this semi-isolated sample was obtained
in a multi-jet enriched sample.

\section{Final Event Selection}

A set of neural networks, one for each tau type, has been developed
to separate the tau leptons from jets. These neural networks make
use of input variables that exploit the tau signature such
as longitudinal and transverse shower shapes and isolation in the
calorimeter and the tracker. The neural network is trained
using tau MC events as signal and multi-jet events from data
as background to produce a variable that peaks near one for
real taus and zero for jets. The tau candidate is required
to have a neural network output greater than 0.9. In the
case of type-3 taus this is tightened to 0.95 due to the
larger multi-jet background.

It is also possible for muons to fake type-1 or type-2 tau candidates.
These fakes are removed by ensuring that type-1 or type-2 tau candidates do not match
to a reconstructed muon within a cone of radius $\Delta R_{\mu\tau}=0.5$.

After selecting events with a high neural network output, there is still
a considerable amount of background from \Wjets ~production.
To remove these events, the reconstructed W boson mass, 
$M_{\rm W}=\sqrt{2 E^{\nu}  E^{\mu}  (1 - \cos \Delta \phi)}$
is used, where 
$E^{\nu} = \MET  p^{\mu} / p_{\rm T}^{\mu}$
is the estimated neutrino energy, calculated using the ratio of
the muon momentum $p^{\mu}$ and muon transverse momentum $p_{\rm T}^{\mu}$.
For real W boson decays, this variable peaks near the W boson mass, whereas
for the signal and the $\Ztt$ background the variable peaks at zero.
Events with $M_{\rm W}>20$~GeV are rejected.

\begin{figure}[htbp]
   \begin{center}
     \includegraphics[height=2.0in]{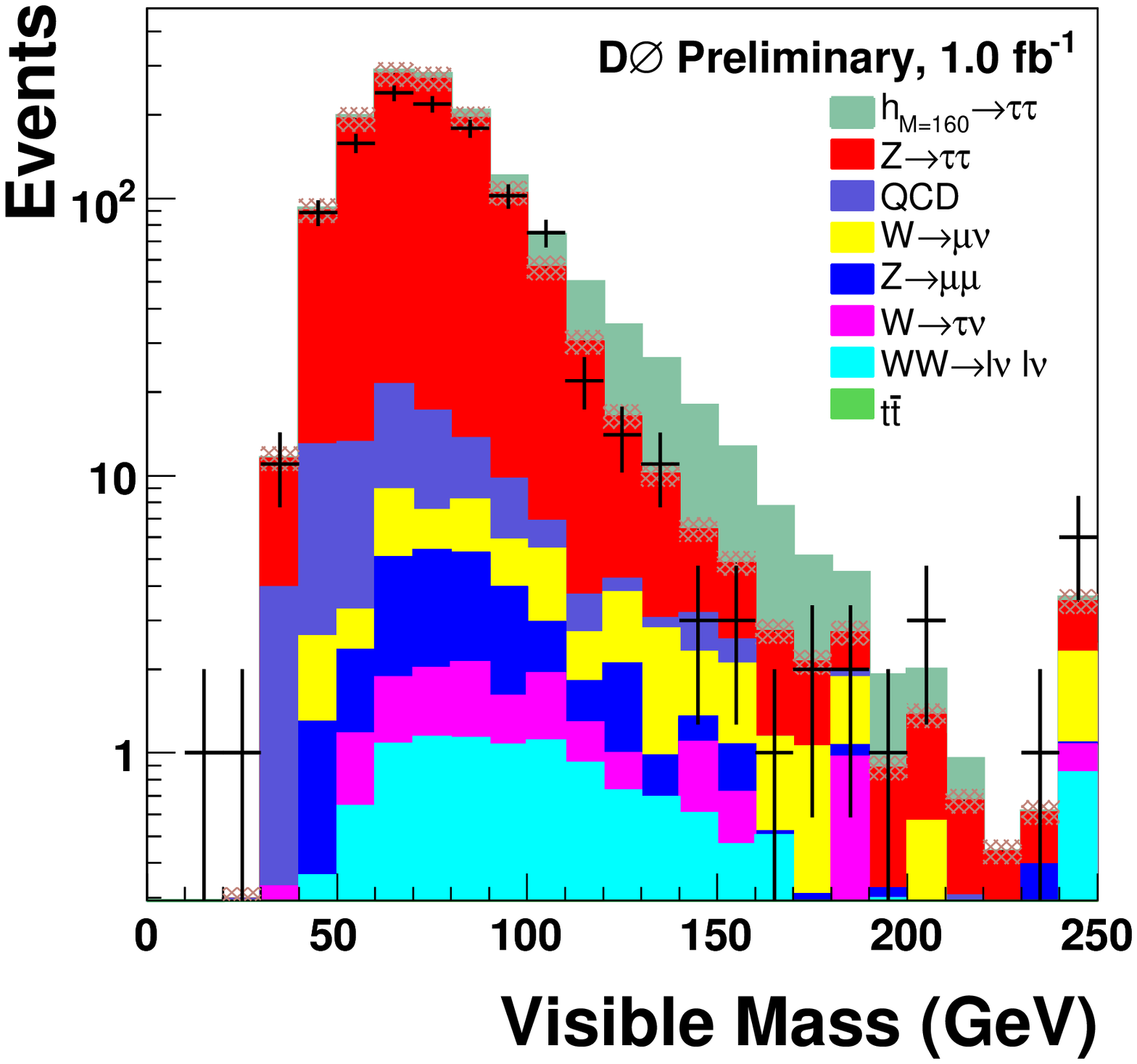}
     \includegraphics[height=2.0in]{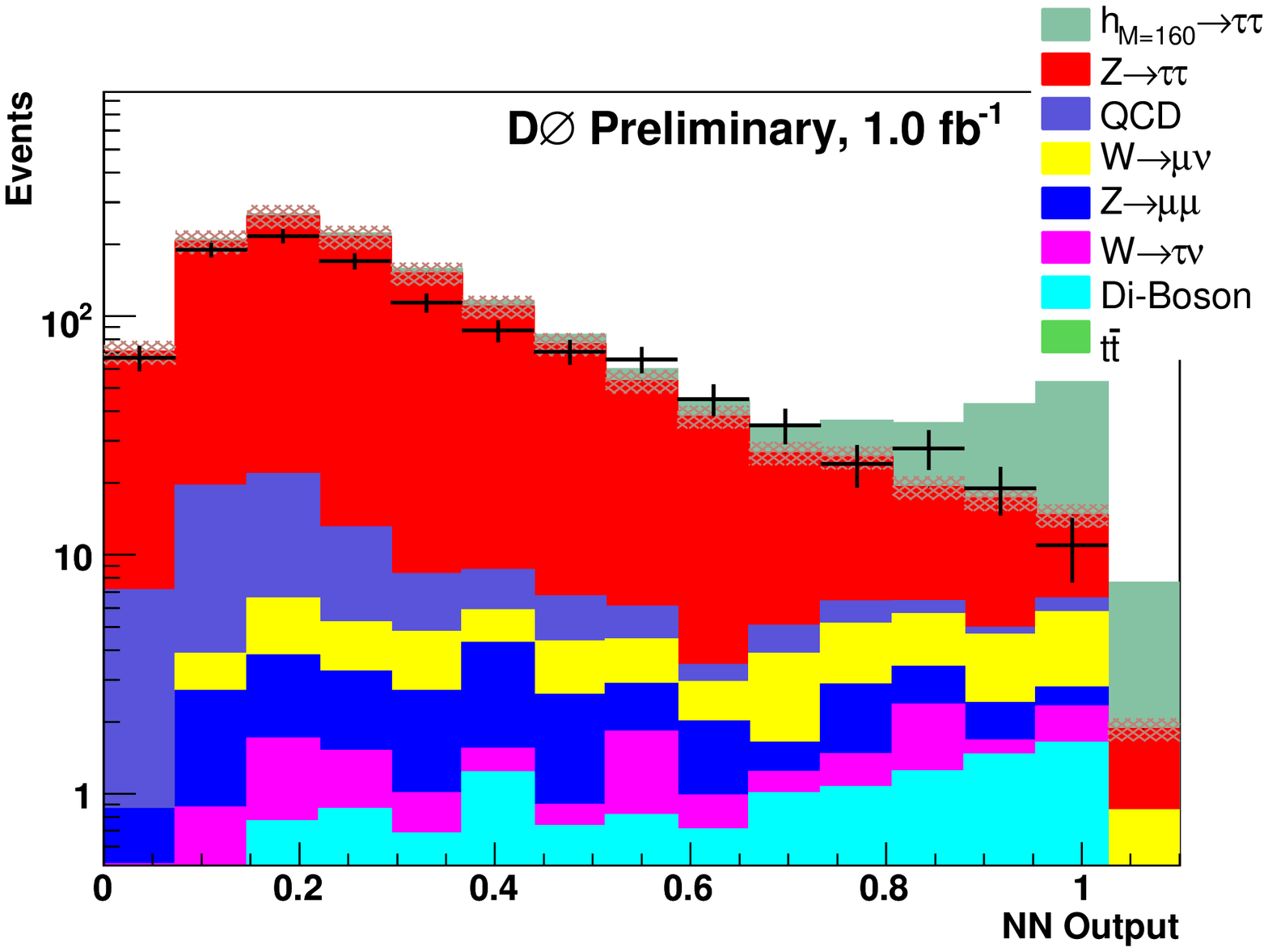}
   \end{center}
   \caption{
     Distribution of a) the visible mass $M_{\rm vis}$ 
     and b) neural network output distribution for a Higgs mass of 160 GeV
     with all selections
     applied. The data, shown with error
     bars, are compared to the sum of the expected backgrounds.
     Overflow events are added to the last bin. Also shown, in light green,
     is the signal for a Higgs mass of $160$~GeV, normalized
     to a cross-section of $10$~pb.
     The systematic uncertainty on the background normalisation
     is $10\%$ and is shown by the shaded area.
     \label{fig-mvis}}
\end{figure}

To achieve the best separation of
the signal from background, neural networks 
were trained for different Higgs mass points 
using kinematical variables.
The distribution of the visible mass $M_{\rm vis}$ and 
the optimised neural networks for a Higgs mass
of $160$~GeV is shown in Figure~\ref{fig-mvis}.
There is good agreement between the background expectation
and the data.

\section{Results and Conclusion}

Limits on the cross-section for
Higgs boson production times the branching fraction into tau leptons
are derived at $95\%$ Confidence Level (CL). 
The output from the neural networks,
shown in Figure~\ref{fig-mvis} for one mass point, 
are used in the limit calculation. The distributions for the three 
tau types are used separately.
The cross-section limits are calculated with the
$\CLS$ method~\cite{bib-limit}. 

\begin{figure}[htbp]
   \begin{center}
     \includegraphics[height=2.0in]{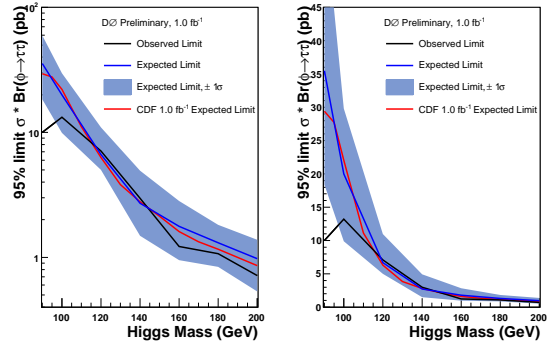}
   \end{center}
\caption{
Observed and expected  $95\%$ Confidence Level upper limit
on the cross-section times branching ratio, using the neural network
shown on both a log scale and a linear scale. The band
represents the $\pm 1\sigma$ uncertainty on the expected limit.
Also shown is the expected limit from the
recent CDF result.
\label{fig-xlimit}
}
\end{figure}

There are various sources of systematic uncertainties that affect signal
and background. The most important are the uncertainty on 
the integrated luminosity ($6.1\%$), the trigger efficiency ($3\%$),
the tau energy scale ($1-11\%$), the uncertainty in the signal acceptance
due to choice of parton distribution function ($3.9-4.6\%$),
the uncertainty of the tau track matching efficiency ($4\%$),
the uncertainty on the tau reconstruction efficiency ($3\%$),
the theoretical uncertainty on the Z cross-section ($5\%$) and the
uncertainty on the modeling of the multi-jet background ($3\%$).
All systematic uncertainties are included in the calculation
of the expected and observed limits,
assuming $100\%$ correlation between signal and background where
appropriate.
The expected
and observed limits are shown in Figure~\ref{fig-xlimit} as a
function of the hypothetical Higgs mass.

\begin{figure}[htbp]
   \begin{center}
     \includegraphics[height=2.0in]{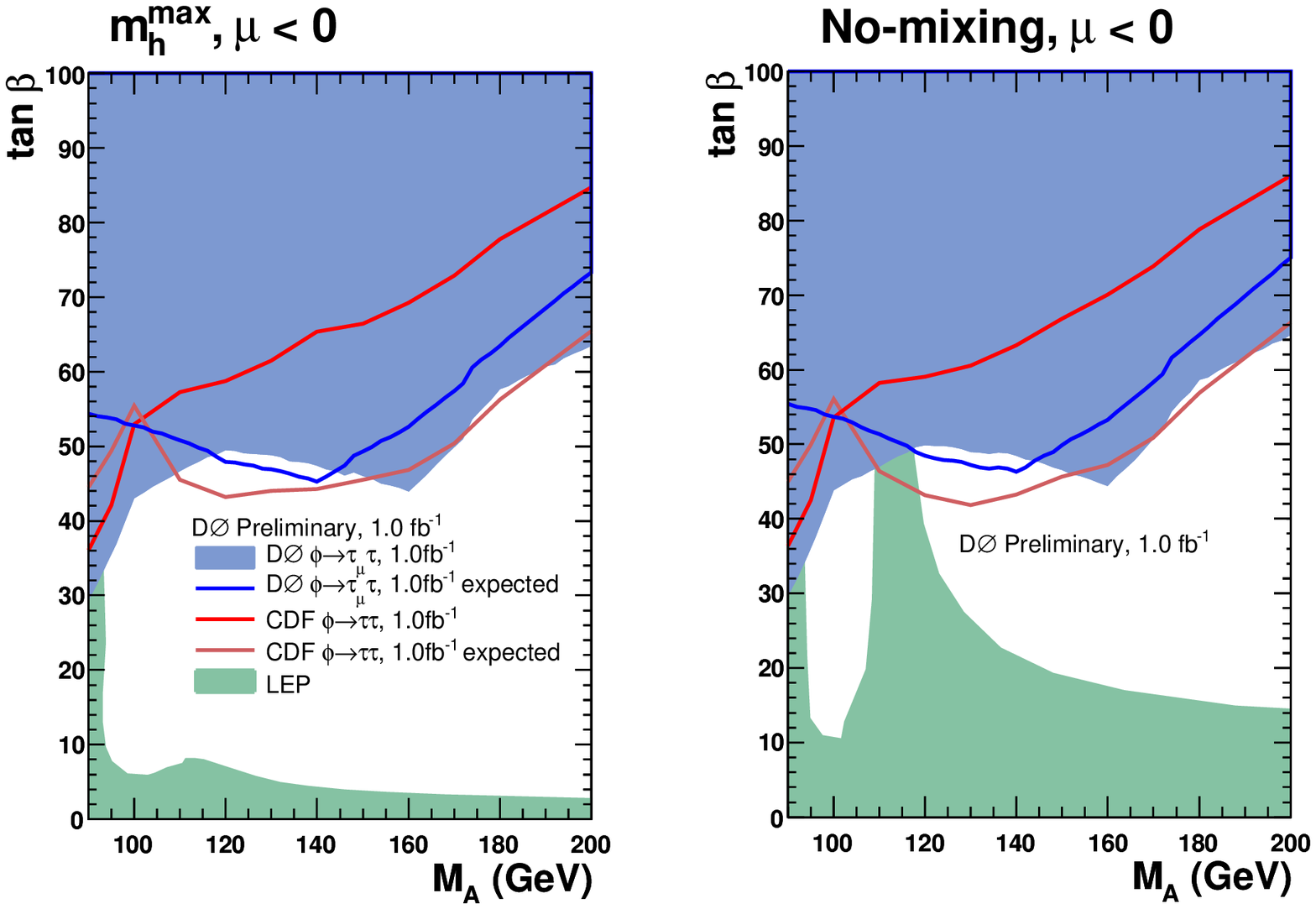}
     \includegraphics[height=2.0in]{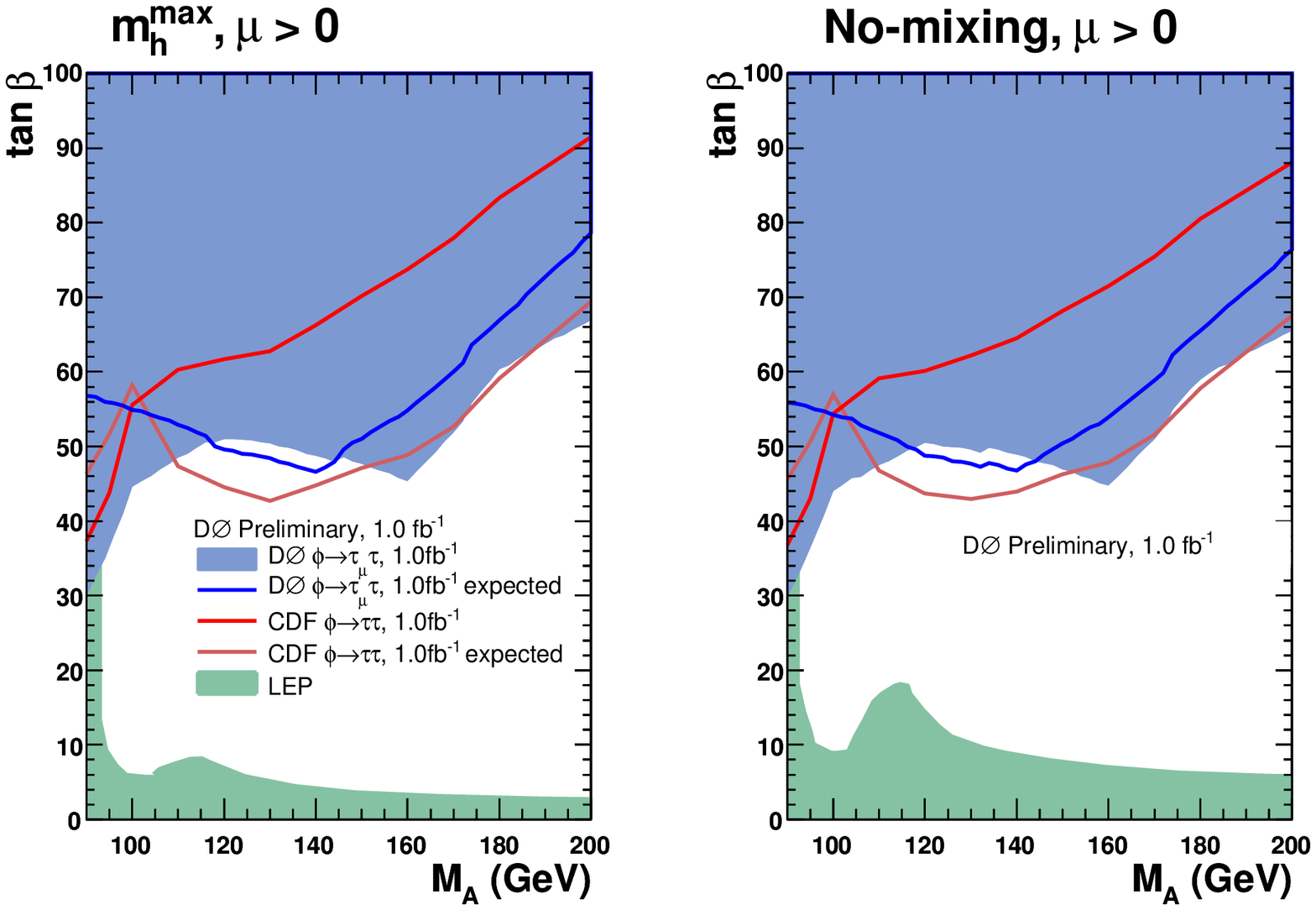}
   \end{center}
\caption{
  Excluded region in the $\tan\beta-m_A$ plane for $\mu < 0$
  in a) the $m_h^{\rm max}$ scenario and b) the no-mixing scenario and
  excluded region in the $\tan\beta-m_A$ plane
  for $\mu > 0$
  in c) the $m_h^{\rm max}$ scenario and d) the no-mixing scenario.
  \label{fig-tanbeta}}

\end{figure}

In the MSSM, the masses and couplings of the Higgs bosons depend
on $\tan\beta$ and $m_A$ at tree level. Radiative corrections
introduce additional dependencies on SUSY parameters. In this analysis, 
the $m_h^{\rm max}$ and no-mixing
scenarios are studied. The corresponding excluded regions
in the $\tan\beta-m_A$ plane are shown in Figure~\ref{fig-tanbeta}.
The cross-section at each $\tan\beta-m_A$ 
point was calculated using FeynHiggs 2.5.1~\cite{bib-feyn}
by adding the $gg\to\phi$ and $\bbbar\to\phi$
cross-sections for a given $m_A$.

In the mass region $90<m_A<200$~GeV,  $\tan\beta$ values larger than 40-60
are excluded for the no-mixing and the $m_h^{\rm max}$ benchmark
scenarios. These results are the most constraining limits
from the Higgs to $\tau^+\tau^-$ decay channel to date.

\end{document}